\documentclass[useAMS,usenatbib]{mnras}
\usepackage{times}
\usepackage{epsfig}
\usepackage{amssymb}
\usepackage{amsmath}
\title[Dust and supernova SN~1995N]
{Evidence for late-time dust formation in the ejecta of supernova SN~1995N from emission-line asymmetries}
\author[R. Wesson et al.]
{R. Wesson$^{1,2}$, A. M. Bevan$^1$, M.J. Barlow$^1$, I. De Looze$^{1,3}$, M. Matsuura$^2$, \newauthor
G. Clayton$^4$, J. Andrews$^5$\\
$^1$Department of Physics and Astronomy, University College London, Gower Street, London WC1E 6BT, United Kingdom\\
$^2$School of Physics and Astronomy, Cardiff University, Queens Buildings, The Parade, Cardiff CF24 3AA, UK\\
$^3$Sterrenkundig Observatorium, Ghent University, Krijgslaan 281-S9, B-9000 Gent, Belgium\\
$^4$Department of Physics and Astronomy, Louisiana State University, Baton Rouge, LA 70803, USA\\
$^3$Gemini Observatory, 670 N. A'ohoku Place, Hilo, HI 96720, USA\\
}
\date{Received:}

\begin{document}
\maketitle

\begin{abstract}

We present a study of the dust associated with the core-collapse supernova SN~1995N. Infrared emission detected 14--15 years after the explosion was previously attributed to thermally echoing circumstellar material associated with the SN progenitor. We argue that this late-time emission is unlikely to be an echo, and is more plausibly explained by newly formed dust in the supernova ejecta, indirectly heated by the interaction between the ejecta and the CSM. Further evidence in support of this scenario comes from emission line profiles in spectra obtained 22 years after the explosion; these are asymmetric, showing greater attenuation on the red wing, consistent with absorption by dust within the expanding ejecta. The spectral energy distribution and emission line profiles at epochs later than $\sim$5000 days are both consistent with the presence of about 0.4~M$_\odot$ of amorphous carbon dust. The onset of dust formation is apparent in archival optical spectra, taken between 700 and 1700 days after the assumed explosion date. As this is considerably later than most other instances where the onset of dust formation has been detected, we argue that the explosion date must be later than previously assumed.

\end{abstract}

\begin{keywords}
Supernovae: general; supernovae: individual: SN 1995N; radiative transfer; (ISM) dust, extinction
\end{keywords}

\section{Introduction}

Understanding the origins of cosmic dust has been of widespread interest since the discovery that galaxies can contain up to 10$^{8}$ M$_\odot$ of dust at redshifts of 6 and beyond (e.g. \citealt{bertoldi2003}). For dust to be produced so early in a galaxy's evolution requires that massive stars play a role, and efficient dust condensation has long been predicted in the expanding ejecta of core collapse supernovae (CCSNe; e.g. \citealt{todini2001,nozawa2003,cherchneff2009}). However, observational estimates of dust masses in supernova remnants have often fallen far short of the predicted amounts, and the amounts required to balance the high-redshift dust budget, leading some authors to argue that CCSNe do not form dust in any cosmologically meaningful quantity (e.g. \citealt{meikle2011}). New observations over the past decade have bolstered the case for supernova dust formation, starting with the discovery of 0.4-0.7~M$_\odot$ of newly formed dust in the remnant of SN~1987A, 23 years after its explosion (\citealt{matsuura2011}), which was confirmed beyond doubt in follow-up observations to be cold dust within the expanding remnant, and not circumstellar material or line emission contaminating estimates of the continuum flux (\citealt{indebetouw2014}). Significant masses of cold dust have also been found in much older remnants including Cas~A (\citealt{barlow2010,arendt2014,delooze2017,niculescuduvaz2021}), the Crab nebula (\citealt{gomez2012,owen2015,delooze2019}), the Sagittarius A East remnant near the Galactic Centre (\citealt{lau2015}), G54.1+0.3 (\citealt{temim2017,rho2018}) and several pulsar wind nebulae (\citealt{rho2018,chawner2019,chawner2020}). How much newly-formed supernova dust will ultimately survive the passage of the reverse shock and reach the interstellar medium remains debated (\citealt{nozawa2007,cherchneff2010,cherchneff2013,kirchschlager2019}).

These dust mass estimates were all made by fitting models to observed spectral energy distributions (SEDs). A major difficulty with this approach is that newly formed dust, if heated only by radioactive decay within the expanding ejecta, cools very rapidly, becoming invisible at near-infrared wavelengths within a few years of supernova explosions. The far-infrared observations necessary to fully constrain the mass of cold dust are impossible to obtain beyond the Local Group, a volume in which only SN~1987A has been observed in the last century.

To reconcile the apparent discrepancy between low dust masses detected in young ($<$3 years old) remnants and high dust masses in older remnants, \citet{gall2014} proposed that the formation of dust could be described by a broken power law, implying slow dust formation at first, accelerating at later times when most supernovae cannot be detected in the mid-IR. \citet{wesson2015} estimated dust masses at five epochs in SN~1987A, fitting them with a sigmoid curve which agreed well with the power law proposed by \citet{gall2014} at the epochs concerned, and predicted a final dust mass of 1.0~M$_\odot$, with the rate of dust formation reaching a maximum some 3000 days after the explosion. In the specific case of interacting supernovae, \citet{sarangi2018} and \citet{sarangi2022} have shown that heating from the radiation produced by ejecta-CSM interaction can delay dust formation and that a relatively high final dust mass is possible. However, this delayed dust formation and high final mass is not currently predicted for supernovae in general: \citet{sarangi2015} predict that dust formation should be complete within two years for an SN~1987A-type supernova, while \citet{sluder2018} predict completion within three years. These authors find total dust masses for an SN~1987A-like progenitor of 0.035 and 0.2~M$_\odot$ respectively - factors of 30 and 5 below the observationally-inferred value.

A new technique for estimating dust masses was presented by \citet{bevan2016}; instead of relying on far-IR observations which are hard or impossible to obtain for most supernovae, they presented {\sc damocles}, a code which calculates the profiles of emission lines arising within the expanding remnant. These emission lines are frequently observable at very late times (\citealt{milisavljevic2012, bevan2017, niculescu-duvaz2022}), and are affected by the presence of dust. Emission from the receding side of the ejecta passes through a column of dust which emission from the near side does not, resulting in an asymmetric profile with an apparent blue shift (\citealt{lucy1989}). Applying this technique to SN~1987A, \citet{bevan2016} found dust masses increasing slowly with time, in good agreement with values derived from SED modelling, supporting the picture of delayed dust formation. \citet{dwek2015} and \citet{dwek2019} have proposed that several tenths of a solar mass of silicate dust could in fact be present at earlier times, if in sufficiently dense clumps, but \citet{wesson2021} investigated a large parameter space and found that at day $\sim$800, the emission line profiles and spectral energy distribution could only be simultaneously reproduced with a mass of around 0.001~M$_\odot$ of amorphous carbon dust.

An ongoing difficulty is the paucity of supernovae observed - with either technique - at epochs of a few thousand days after discovery. To fill in this observational gap and build up a clearer picture of the evolution of dust in supernova remnants, we have carried out a programme of spectroscopic observations of remnants older than $\sim$8 years. In this paper, we present a study of one such remnant, that of SN~1995N. This Type IIn supernova was discovered some time after exploding (\citealt{pollas1995}), and infrared detections 14--15 years after its explosion have been attributed to thermal echoes from a dusty circumstellar medium (\citealt{vandyk2013}). We present new radiative transfer models which show that in fact this emission cannot be a thermal echo, but rather arises from newly formed dust in the supernova ejecta, heated by the interaction of the ejecta with the CSM. The onset of dust formation is apparent in archival spectra taken 500-700 days after the explosion, while at epochs of 5500-7700 days, emission line profiles are consistent with the presence of 0.4~M$_\odot$ of dust in the expanding ejecta, as derived from the SED.

\section{Observations}
\label{observations}

SN~1995N was discovered on 1995 May 5 (\citealt{pollas1995}). It occurred in the galaxy pair MCG-02-38-017 (Arp 261), which are undergoing major interaction. They are at a distance of $\sim$24~Mpc and have a redshift of 0.006170 (\citealt{meyer2004}). The SN has been studied across the electromagnetic spectrum, being one of the first young supernova remnants to be detected at X-ray wavelengths (\citealt{fox2000}), as well as being observed on numerous occasions in the optical (\citealt{fransson2002}). It has also been observed with \textit{Spitzer} and \textit{WISE} at infrared wavelengths (\citealt{vandyk2013}), and at radio wavelengths with the Very Large Array and other facilities (\citealt{chandra2009}).

\subsection{Explosion date}
\label{epoch}

Before proceeding, we consider the likely explosion date of SN~1995N. In reporting its discovery, \citet{pollas1995} noted that its H$\alpha$ emission line profile was symmetric, with a full width at half maximum of 34{\AA}. The implied expansion velocity is therefore $\sim$1500\,kms$^{-1}$. \citet{pollas1995} reported that this was similar to the H$\alpha$ profile in SN~1993N, 10 months after its explosion. This has been taken by subsequent authors to imply an explosion date around July 1994, or earlier. Neither the discovery spectrum of SN~1995N nor the spectrum of SN~1993N are available via either the Open Supernova Database (\citealt{guillochon2017}) or the Weizmann Interactive Supernova Data Repository (WISeREP; \citealt{yaron2012}). However, this intermediate-width H$\alpha$ emission may arise either from the ejecta, or from interaction with circumstellar material. In either case, there is no reason to expect a dependence of the line width on the age of the supernova. In the latter case, the geometry and location of the circumstellar material would determine the line profile. Considering the former case, in their extensive compilation of the spectra of Type~{\sc ii} supernovae, \citet{gutierrez2017} find a large range in expansion velocities from H$\alpha$, with values between 1500 and 9600\,kms$^{-1}$ at 50 days post-explosion. As the width of H$\alpha$ is not a strong indicator of age, we therefore considered other possible constraints on the explosion date.

\begin{figure}
\includegraphics[width=0.5\textwidth]{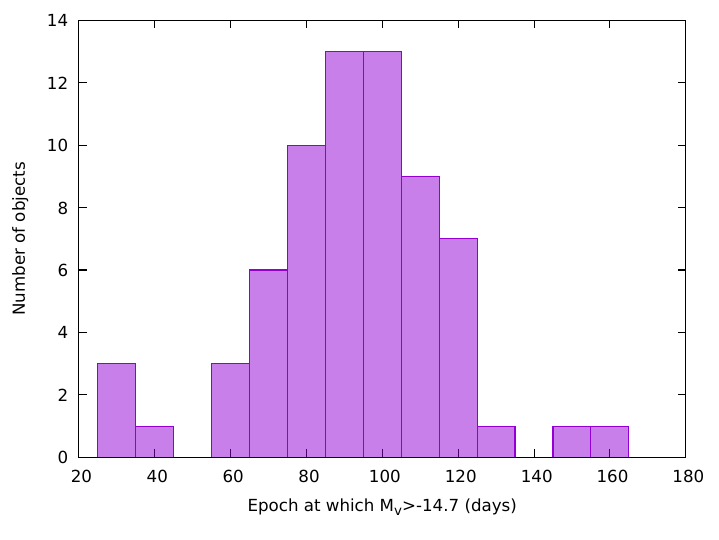}
\caption{For 68 core-collapse supernovae, the distribution of epochs after discovery at which the object's absolute magnitude became fainter than that of SN~1995N at the time of its discovery. The light curves used to create this histogram are from \citet{anderson2014}.}
\label{declineepochs}
\end{figure}

When discovered, the supernova had m$_V$=17.5. The distance modulus to Arp 261 is 31.9 (\citealt{meyer2004}), and extinction towards the supernova was reported to be low by \citet{fransson2002}, who gave E(B-V)=0.11, corresponding to A$_V$=0.34. The absolute magnitude of the supernova at the time of discovery was then M$_V$=-14.7. \citet{nyholm2020} studied the light curves of Type~IIn supernovae and determined an average decline rate over the first few tens of days of 0.027\,mag\,d$^{-1}$. However, their compilation of light curves included very few objects with data at epochs of 100 days or older. \citet{anderson2014} compiled the light curves of 116 Type~{\sc ii} supernovae, of which 68 objects had data at epochs later than 100 days. For these objects, we determined the epoch at which the supernova became fainter than M$_V$=-14.7. The distribution of values is strongly peaked at around 100 days; the median value was 98.4 days (Figure~\ref{declineepochs}). \citet{anderson2014} find a decline rate in the initial phases of 0.0265 mag\,d$^{-1}$ for their sample. The magnitude at discovery of SN1995N is thus consistent with an age of $\sim$100 days. An older age is not ruled out but would require that its decline rate was slower than average.

If the supernova did in fact explode in July 1994 or earlier, then it would have been accessible at the beginning of the night to observatories at latitudes southward of about 30$^{\circ}$N (further north, its airmass at the end of astronomical twilight would have been $\gtrsim$2). If its peak magnitude was 2 magnitudes brighter than its discovery magnitude, it would have had m$_V$=15.5, significantly brighter than the median discovery magnitude of 17.5 for the 41 supernovae discovered in 1994. However, although several systematic supernova searches were underway in the 1990s, including some at well-placed observatories (\citealt{cappellaro1997}), discoveries were still clearly incomplete at this brightness, with an average of 9.7 discovered annually at brighter than m$_v$=15.5, compared to an average of 23.4 per year in the 2000s.\footnote{according to the list of supernovae up to 2015 compiled by CBAT, \url{http://www.cbat.eps.harvard.edu/lists/Supernovae.html}} Therefore, there is no clear upper limit on the supernova's age at the time of discovery based on its light curve. However, as discussed in Section~\ref{earlyspectra}, spectra taken 700 days after the nominal explosion date of 1994 July 4 show a very symmetric H$\alpha$ line profile, which becomes increasingly extinguished on its redward side at days 1000 and 1700. The symmetry of the profile constrains the amount of dust in the ejecta by this epoch to be less than about 4$\times$10$^{-5}$~M$_\odot$ of amorphous carbon dust. In supernovae where dust formation in the ejecta is detected, the onset of the formation is typically observed after 300-500 days; in SN~1987A, an infrared excess began to develop around 400 days after its explosion (\citealt{wooden1993}). No dust formation until at least 700 days after the explosion would be a unusually late epoch, especially considering that supernovae with a dense CSM often show much earlier dust formation (\citealt{smith2008, andrews2016}).

We therefore infer that at the time of its discovery, the supernova was not as old as previously assumed, and adopt an explosion epoch of 1995 January 25, 100 days before it was discovered. While the constraints we have considered do not definitively rule out an earlier explosion, the age of 10 months at the time of discovery that has been quoted in several papers is not well supported and is likely to be an overestimate.

\subsection{Archival infrared photometry}
\label{spitzerwise}

\begin{figure}
	\includegraphics[width=0.5\textwidth]{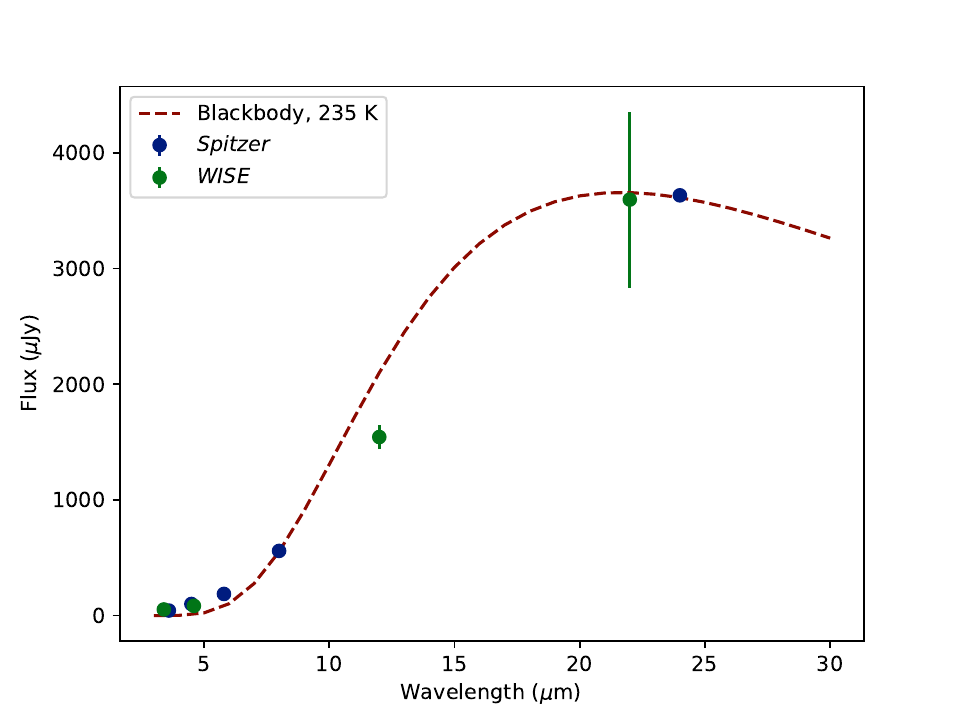}
	\caption{Combined \textit{Spitzer}+\textit{WISE} SED of SN~1995N, with an average epoch of 5366 days. Also plotted is a blackbody spectrum with a temperature of 235\,K, optimised to fit datapoints with $\lambda \geq$8$\mu$m.}
	\label{spitzerwisesed}
\end{figure}

SN~1995N was observed with the \textit{Spitzer Space Telescope} as part of program 50454, with the Infrared Array Camera (IRAC; \citealt{fazio2004}) on 19 March 2009 (day 5167), and with the Multiband Imaging Photometer for Spitzer (MIPS; \citealt{rieke2004}) on 29 March 2009 (day 5177). These observations covered wavelengths of 2.4, 4.5, 5.6 and 8~$\mu$m (IRAC), and 24 and 70~$\mu$m (MIPS). The supernova was detected at all wavelengths except 70~$\mu$m, where only a weak upper limit to its flux was obtained. The supernova was also observed at wavelengths of 3.4, 4.6, 12 and 22 $\mu$m with the \textit{Widefield Infrared Survey Explorer} (\textit{WISE}; \citealt{wright2010}); these observations took place on 2010 January 01 (5455 days) and 2010 July 31 (5666 days). \citet{vandyk2013} calculated colour-corrected flux densities from these observations, which we use for our radiative transfer modelling. These fluxes are reproduced in Table~\ref{irfluxes}. Figure~\ref{spitzerwisesed} shows the combined \textit{Spitzer+WISE} SED, with a blackbody spectrum derived by fitting the datapoints with $\lambda \geq$8$\mu$m; this fit yields a temperature of 235\,K, a radius of 1.6$\times$10$^{17}$cm (0.052 pc), and a luminosity of 1.8$\times$10$^{40}$erg\,s$^{-1}$.

\begin{table}
\begin{tabular}{llll}
\hline
Date & Epoch (days) & Wavelength ($\mu$m) & Flux density ($\mu$Jy) \\
\hline
19 March 2009 & 5167 & 3.6 & 43.8 $\pm$ 1.1\\
19 March 2009 & 5167 & 4.5 & 100.4 $\pm$ 1.7 \\
19 March 2009 & 5167 & 5.8 & 187.1 $\pm$ 7.1 \\
19 March 2009 & 5167 & 8.0 & 559.0 $\pm$ 8.9 \\
29 March 2009 & 5177 & 16 & 3631.6 $\pm$ 36.3 \\
29 March 2009 & 5177 & 70 & $<$3$\times$10$^{4}$ \\
01 January 2010 & 5455 & 3.4 & 53 $\pm$ 6 \\
01 January 2010 & 5455 & 4.6 & 84 $\pm$ 12 \\
01 January 2010 & 5455 & 12 & 1543 $\pm$ 105 \\
01 January 2010 & 5455 & 22 & 3595 $\pm$ 760 \\
\hline
\end{tabular}
\caption{SN~1995N infrared fluxes from 2009-2010 \textit{Spitzer} and \textit{WISE} observations, as reported by \citet{vandyk2013}. Epochs refer to our assumed explosion date of 1995 January 25.}
\label{irfluxes}
\end{table}

\subsection{New and archival optical spectroscopy}

We observed SN~1995N with X-shooter (\citealt{vernet2011}) on the Very Large Telescope at Paranal, Chile, as part of program 097.D-0525(A) (PI: Barlow). The object was observed four times in a period of 8 nights, on 24/25 Jul 2016, 28/29 Jul 2016, 31 Jul/1 Aug 2016 and 1/2 Aug 2016. We observed using the instrument in its integral field unit (IFU) mode, to ensure that the supernova flux caught was maximised regardless of any uncertainty in the position of the object and the pointing of the instrument (the object was acquired using a blind offset from a nearby 19th magnitude star). The resolution of the instrument in this mode is $\lambda$/$\Delta\lambda$=8600, 13500 and 8300 for the UVB, VIS and NIR arms respectively, and the field of view is 4$\times$1.8 arcsec. Each observation consisted of 2$\times$1146s exposures in the UVB arm, 2$\times$ 1052s exposures in the VIS arm, and 12$\times$ 200s exposures in the NIR arm. The data were reduced using Reflex (\citealt{freudling2013}) to run version 2.6.8 of the X-shooter pipeline. For IFU staring data, this does not include cosmic ray removal, so we performed that separately using a python implementation of the LACosmic algorithm of \citet{vandokkum2001}, originally written by Malte Tewes.

The spectra of 28 and 31 July and 1 August had very similar continuum levels, while on 24 July, when clouds were present at Paranal and the seeing was noted as varying, the spectra show a much lower continuum. We attributed that to the poor observing conditions and did not use data from this night in our subsequent analysis. The mean date of the three observations used is 2016 July 31, corresponding to 7858 days after the assumed explosion date of 1995 January 25. We extracted the supernova spectrum from a circular aperture 1 arcsec in diameter. For each spectrum, we converted the observed wavelengths to the heliocentric rest frame, removing the geocentric velocity using {\sc rv} (\citealt{wallace2014}), and using the redshift of 0.006170 for the host galaxy (\citealt{meyer2004}). The spectra were then combined by taking the median of the three fluxes at each wavelength.

\begin{figure*}
\includegraphics[width=\textwidth]{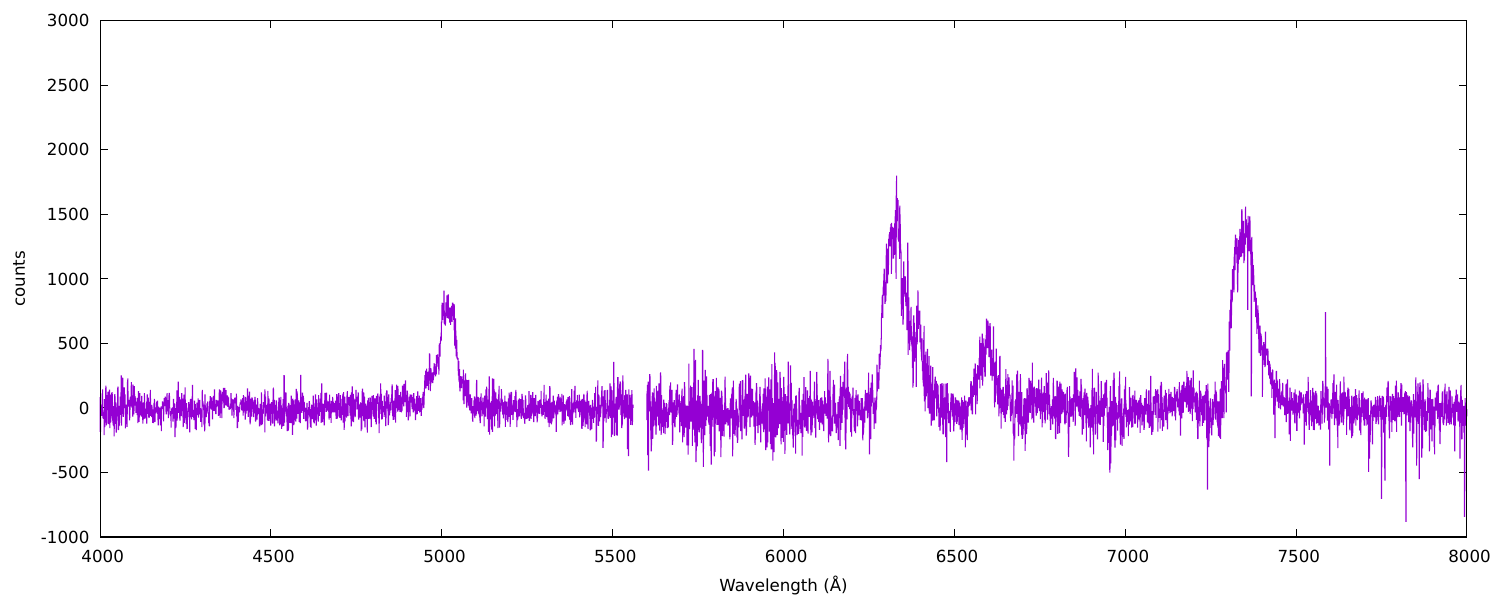}
\caption{An extract of the X-shooter spectrum of SN~1995N taken in July 2016, showing the broad and asymmetric emission line profiles of H$\alpha$, [O~{\sc i}] (6300,6363{\AA}), [O~{\sc ii}] (7320,7330{\AA}) and [O~{\sc iii}] (4959,5007{\AA}).}
\label{snspectrum}
\end{figure*}

At 21-22 years after its explosion, the object is faint, but we clearly detect broad emission from [O~{\sc i}] 6300,6363{\AA}, [O~{\sc ii}] 7320,72330{\AA} and [O~{\sc iii}] 4959,5007{\AA}. H$\alpha$ is also weakly detected. The faintness of the object meant that the line profiles were obscured by many weak sky lines. The strength of these lines relative to the source means that the Poisson noise and hence residuals are always large relative to the strength of the SN emission when fitting and subtracting them; this was evident when we tried to fit and subtract the lines using both ESO's {\sc skycorr} (\citealt{noll2014}), which calculates an atmospheric emission model, and with {\sc alfa} (\citealt{wesson2016a}), which fits sky lines without reference to a physical model but instead by optimising Gaussian parameters for each line independently. We therefore instead simply masked out sky emission, removing pixels within 2 FWHM of sky lines in the catalogue of \citet{hanuschik2003}. Our July-August 2016 spectrum of the supernova is shown in Figure~\ref{snspectrum}.

SN~1995N was observed with X-shooter on 2010 April 15 and 2010 July 13 (5559 and 5648 days after the 1995 January 25 explosion epoch), as part of ESO program 084.D-0265 (PI Benetti). We downloaded these data from the ESO archive, and reduced them using Reflex as with our new data. These data show many of the broad profiles still visible in 2016, with brighter broad emission at H$\alpha$.


\section{Arguments against a light echo as the source of late-time IR emission}
\label{csm}

\citet{vandyk2013} calculated SEDs from idealised dust clouds, and fitted the observations of SN~1995N with either 0.05~M$_\odot$ of silicate dust, or 0.12~M$_\odot$ of graphite dust, using optical constants from \citet{draine1984} and \citet{weingartner2001} respectively, using a grain radius of 0.1$\mu$m in both cases. Based on this, he proposed that SN~1995N was surrounded by a dusty circumstellar medium, which was illuminated by the supernova explosion giving rise to a thermal echo detected with various infrared instruments.

\citet{vandyk2013} noted that the \textit{WISE} fluxes agree better with the SED calculated for graphite than that for silicate dust. Similar results have been found for several other supernovae observed in the mid-infrared (\citealt{williams2015}). For SN~1995N, the \textit{WISE} 12~$\mu$m flux is 1.5$\pm$0.1\,mJy, where the silicate model would predict a flux of 4.0\,mJy. Using a non-linear least-squares algorithm to optimise the parameters of these idealised dust clouds, we find that a silicate model with 0.061$M_\odot$ of dust at a temperature of 225\,K gives a fit with $\chi^2$=6035, while a graphite model with 0.113\,M$_\odot$ of dust at 236\,K gives $\chi^2$=3824.

These very large $\chi^2$ values are driven by the shortest wavelength points, for which the fluxes are very low. Restricting the optimisation to datapoints with $\lambda \geq $8$\mu$m, the best fits are achieved with 0.065\,M$_\odot$ of silicate dust at 221\,K ($\chi^2$=400) or 0.166\,M$_\odot$ of graphite dust at 216\,K ($\chi^2$=0.1). Allowing the grain size to vary, we find that 0.16\,M$_\odot$ of graphite dust at T$\sim$220\,K gives good fits for all grain sizes $\leq$0.1$\mu$m, while for silicate dust, extremely large grains somewhat improve the fit: 0.097\,M$_\odot$ of 10$\mu$m silicate grains at T=248\,K gives $\chi^2$=22. Such large grains, if present at late times, would produce a strong red scattering wing in late-time emission line profiles, in contrast to the observations presented in Section~\ref{latespectra}.

\begin{figure}
\includegraphics[width=0.5\textwidth]{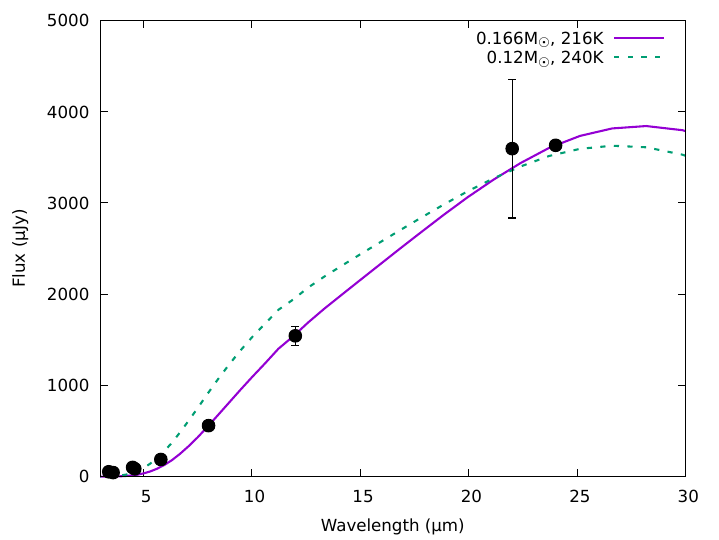}
\caption{The predicted SEDs of idealised graphite dust clouds with the observed fluxes of SN~1995N at days 5167-5455. The dashed curve is from \citet{vandyk2013}, while the solid curve is optimised to fit data points with $\lambda\geq8\mu$m.}
\label{newsimplefit}
\end{figure}

Given the far better fit of the graphite models and the absence of any red scattering wing that might support the presence of very large silicate grains in our late-time spectra, we consider that silicate dust is ruled out, and that under the assumption of an optically-thin point source, a somewhat larger and cooler mass of graphite dust than proposed by \citet{vandyk2013} is necessary to fit the observed SED. The best-fitting graphite SED for a grain size of 0.1$\mu$m is shown in Figure~\ref{newsimplefit}.

\citet{vandyk2013} attributed the emission to flash-heated dust in the circumstellar medium of the progenitor. However, we argue that a thermal echo cannot explain the SED in 2009-2010, because the echoing medium would need to be too far from the SN and too massive to plausibly be circumstellar material.

\begin{figure}
\includegraphics[width=0.5\textwidth]{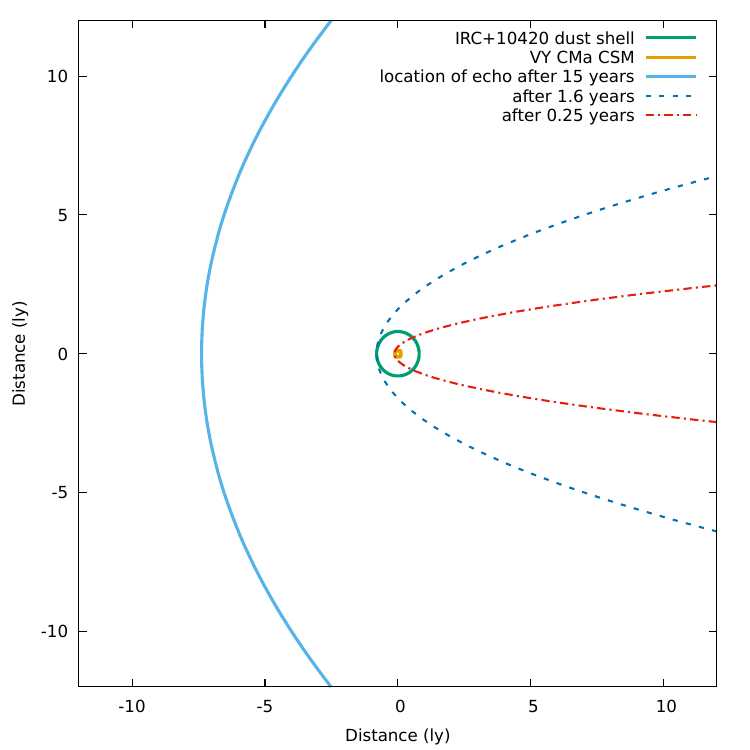}
\caption{Illustration of the propagation of the supernova flash from SN~1995N. This simplified view has the observer towards the right, and shows the extent of the CSM around the extreme evolved stars VY~CMA and IRC~+10420.}
\label{echofigure}
\end{figure}

Dust that has just been heated by emission from a supernova, as seen from Earth, must lie on a surface defined by equal SN-point-Earth light travel times. This surface will be an ellipsoid under the assumption that each SN photon is scattered once only. A light pulse lasting 100 days would then result in emission from a thin region of an extended circumstellar shell lying between two ellipsoidal surfaces. This emitting region will contain only a small fraction of the total mass of the CSM. The echoing dust closest to the supernova will lie directly behind it as seen from Earth. 15 years after the explosion, this nearest echoing dust will lie 7.5 light years (7.1$\times$10$^{18}$ cm) away from the supernova. In comparison, the known circumstellar media of supernovae and supernova progenitors are much smaller. The extreme red supergiant VY~CMa has CSM visible out to about 7750 AU (1.15$\times$10$^{17}$\,cm; \citealt{smith2001a,scicluna2015}), containing up to 10$^{-2}$~M$_\odot$ of dust. The cool hypergiants IRC~+10420 and HD~179821, meanwhile, both have dust shells extending to about 0.25~pc (7.7$\times$10$^{17}$\,cm; \citealt{kastner1995}). The thermal echo from material at these distances would be visible after $\sim$90 and $\sim$600 days respectively. Thus, for material around SN~1995N to be thermally echoing after 15 years, the progenitor would have needed a circumstellar medium far denser and more extensive than those of VY~CMa, IRC~+10420 and HD~179821. Figure~\ref{echofigure} shows a projected 2D view of the echo geometry, indicating the approximate extent of the circumstellar medium of VY~CMa as described by \citet{smith2001}, and the dust shells around IRC~+10420 and HD~179821 as described by \citet{kastner1995}. Three ellipsoids indicate the outer edge of the echoing region after 0.25, 2.0 and 14.8 years respectively.

\begin{figure}
\includegraphics[width=0.5\textwidth]{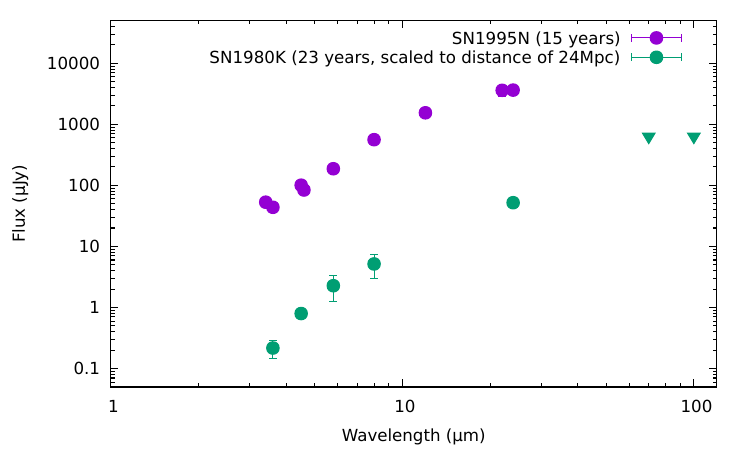}
\caption{Comparison of the mid-IR SEDs of SN~1995N after $\sim$15 years (purple) and SN~1980K after 23 years (green). The fluxes of SN~1980K have been scaled from a distance of 7.7~Mpc (\citealt{eldridge2019}) to the 24~Mpc distance of SN~1995N.}
\label{sn1980k_comparison}
\end{figure}

\citet{sugerman2012} attributed a long-lasting echo associated with SN~1980K to 0.02~M$_\odot$ of carbon-rich dust, in a shell with a radius of about 14 light years, representing a wind-blown shell around the progenitor. They estimated dust temperatures of 200--245\,K, similar to those we estimate for SN1995N. Figure~\ref{sn1980k_comparison} shows the observed SED of SN~1995N, together with SN~1980K's mid-infrared SED scaled to the same distance; this shows that the mid-IR emission from SN~1995N is about 100 times brighter than that of SN~1980K, indicating that if a similar echo were to be responsible, a much higher dust mass and/or a much greater energy input from the supernova would be required. Following the analytical estimate of equilibrium grain temperatures described by \citet{fox2010}, we find that to heat graphite dust grains with a radius of 0.5$\mu$m at a distance from the supernova of 7.5~ly to a temperature of 220\,K would require a supernova luminosity of $\sim$8$\times$10$^{11}$L$_\odot$.

\begin{figure}
\includegraphics[width=0.5\textwidth]{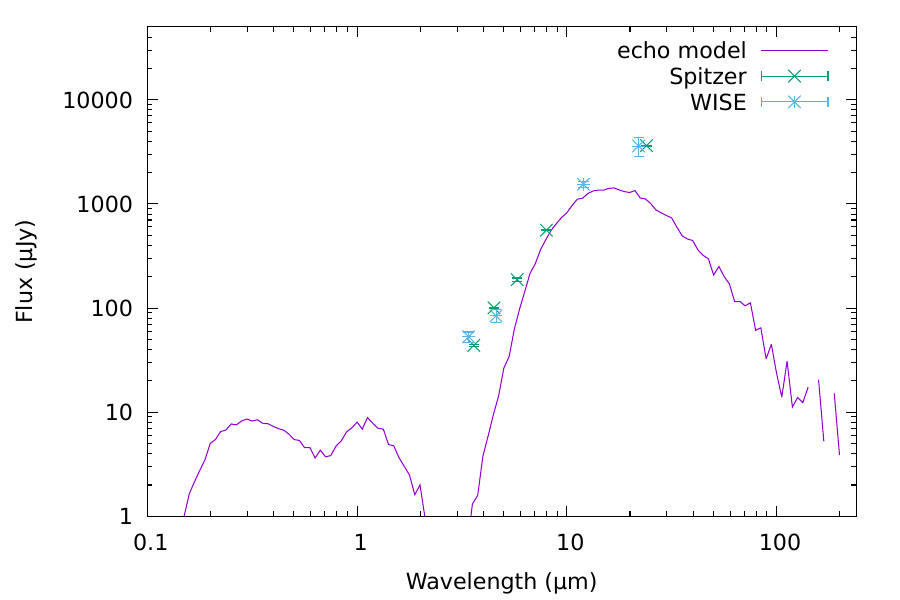}
\caption{The calculated SED of a light echo arising from a shell extending from 2 to 20 ly from the SN, and consisting of amorphous carbon dust grains with radii of 0.1~$\mu$m and a dust mass of 10~M$_\odot$.}
\label{implausibleecho}
\end{figure}

To further demonstrate that the mid-IR emission is unlikely to be a thermal echo, we constructed models using the 3-dimensional radiative transfer code {\sc mocassin}, version 2.02.73 (\citealt{ercolano2003, ercolano2005,ercolano2008}; this release incorporates some computational improvements which significantly reduce the run time of simulations compared to earlier versions. We constructed models to simulate the emission from a dusty sphere extending from 2 to 20 light years from the SN, containing masses of dust up to 10~M$_\odot$, and illuminated by a 100-day pulse. The early photometric evolution of SN~1995N is unknown, but the bolometric light curve of SN~1987A, integrated over its first 100 days, gives an average luminosity of $\sim$4.3$\times$10$^{40}$ erg\,s$^{-1}$ (\citealt{suntzeff1990}). Assuming a luminosity 100$\times$ greater for SN1995N, to maximise the possible echoing flux, we find that to come close to reproducing the observed mid-IR fluxes, dust shells with masses of at least 1~M$_\odot$ are required. Such models are clearly not compatible with the dust being circumstellar in origin, as they would require a CSM mass of upwards of 100$M_\odot$ assuming a gas to dust ratio of 100. Indeed, only models with 10~M$_\odot$ of dust can give fluxes approaching the 3.5\,mJy Spitzer 24~$\mu$m flux; Figure~\ref{implausibleecho} shows a model with 10~M$_\odot$ of amorphous carbon dust grains with radii of 0.1~$\mu$m.

Finally, we note that archival X-shooter spectra taken in 2010 reveal broad emission lines showing red-blue asymmetries. These are discussed further in Section~\ref{latespectra}. Dust which is thermally echoing is exterior to the illuminating source, and will therefore absorb emission equally from the approaching and receding sides of the ejecta. Asymmetric emission lines thus argue strongly in favour of the dust being present within the expanding supernova ejecta.

\section{Ejecta SED modelling}
\label{newdust}

\begin{figure}
	\includegraphics[width=0.45\textwidth]{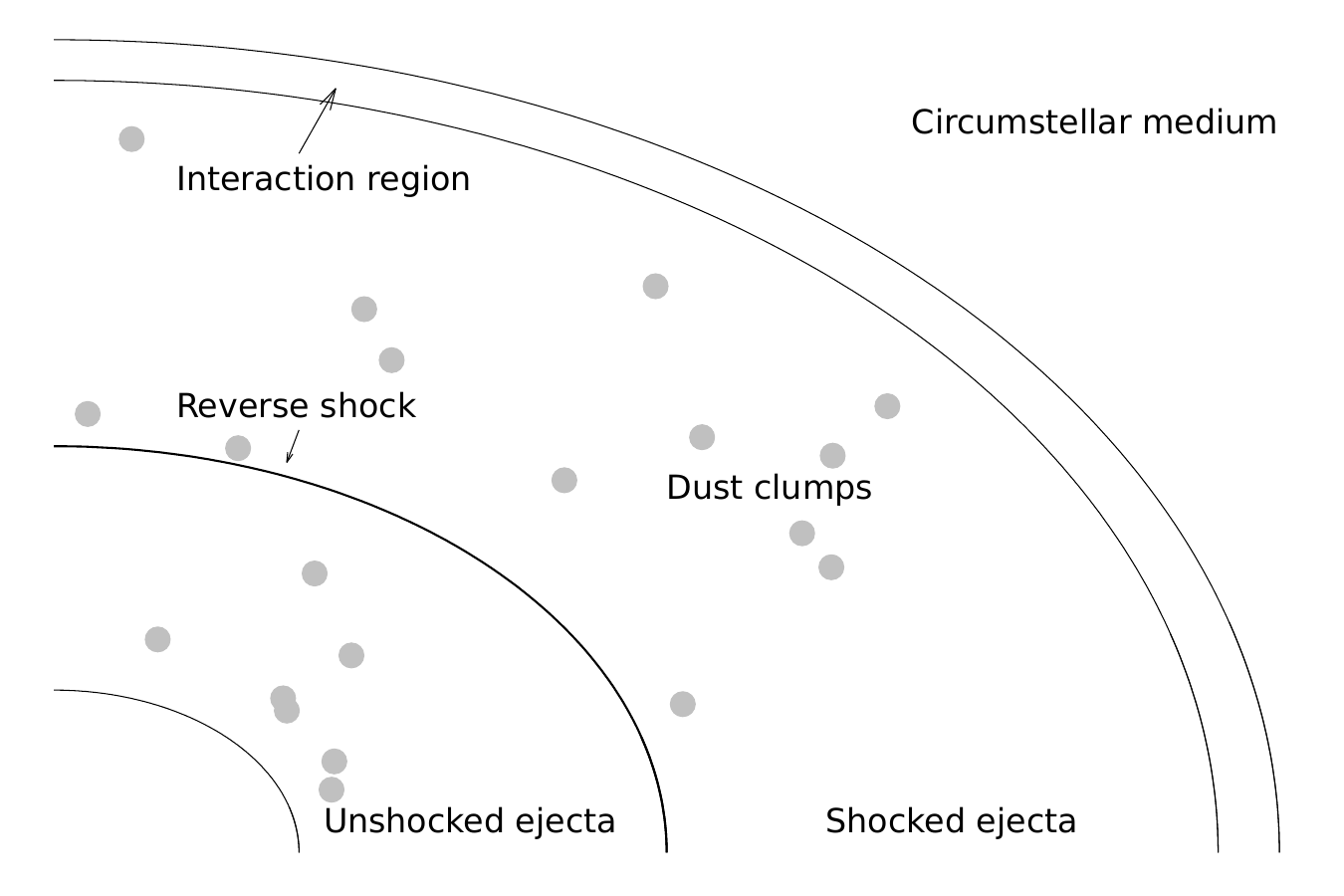}
	\caption{Illustration of the assumed geometry. Any dust in the CSM beyond the forward shock is assumed to have been destroyed by the supernova flash. New dust then forms inside the expanding ejecta and/or in the region between the forward and reverse shocks, and is heated by radiation from the interaction region.}
	\label{schematic}
\end{figure}

We thus consider that an echo model is unviable, and a different heating source is required. \citet{bevan2020} published a series of models of SN~2010jl, in which the interaction of supernova ejecta with a dense CSM was assumed to heat dust both interior and exterior to the interaction region, interior dust being newly formed in the supernova ejecta, and the exterior dust being preexisting circumstellar material. In the case of SN~2010jl, the preexisting CSM was a torus at a distance of 0.3~pc from the supernova, which gave rise to an early thermal echo (\citealt{andrews2011a}). We apply a similar model to SN~1995N, with dust interior to the forward shock being heated by the ejecta-CSM interaction. Evidence for dust forming in the ejecta comes from the developing asymmetry of the H$\alpha$ line reported by \citet{fransson2002}; we quantify this in Section~\ref{earlyspectra}. Figure~\ref{schematic}, based on Figure~4 of \citet{bevan2020}, illustrates the assumed geometry.

We construct 3D radiative transfer models of the ejecta using {\sc mocassin} (\citealt{ercolano2003, ercolano2005}). We adopt a model setup in which all of the dust in the supernova ejecta is contained within clumps, which occupy a fraction of the total volume defined by a volume filling factor, $f$. The collision of the ejecta with circumstellar material is assumed to give rise to an outer shock at a distance of 2.3$\times$10$^{17}$cm from the supernova, corresponding to the distance reached after 5366 days by material travelling at 5,000~km\,s$^{-1}$; \cite{fransson2002} estimated ejecta velocities of 2500-5000\,kms$^{-1}$ from the intermediate-width component of oxygen lines at epochs up to $\sim$1500 days. We assume that the clumps in the expanding ejecta have radii of R$_{\rm out}$/30, based on the theoretical expectation that Rayleigh-Taylor instabilities with characteristic sizes of this order develop at the onset of the supernova explosion (\citealt{arnett1989b}).

New dust within the ejecta may have formed in the unshocked ejecta which has not yet been affected by the reverse shock, or in the cool dense shell between the forward and reverse shocks (\citealt{sarangi2022}). Given the lack of observational constraints on the exact location of the reverse shock and the density profile of the ejecta in the intershock region, we adopt a simplified geometry in which the inner radius of the ejecta is set to 0.2$\times$ the outer radius, and the density is proportional to r$^{-2}$. The ejecta-CSM interaction provides the copious X-rays observed from the source, which then heat the newly formed ejecta dust interior to the interaction region. Based on the SEDs of idealised dust clouds discussed in the previous section, we assume that the dust is pure amorphous carbon, and use optical constants from \citet{zubko1996} (their ``BE" sample, produced by burning benzene in air).

We first modelled the emission by constructing a model in which the heating source was distributed around the surface of a sphere representing the interaction region. However, we found that this could not heat enough of the dust to the observed temperatures. We therefore follow the approach of \cite{bevan2020} by using a diffuse heating source which is colocated with the dust. In this scenario, the assumption is that the dust is being heated by the gas within the expanding ejecta, which is itself heated by the ejecta-CSM interaction. The luminosity of this heating source is constrained by the integrated emission in the infrared, although the luminosity of the forward shock itself would be expected to be much higher.

The parameter space that we investigated covered clump volume filling factors between 0.02 and 0.5; grain sizes from 0.1 to 5.0 $\mu$m; and dust masses between 10$^{-3}$ and 1.0~M$_\odot$. The SED is largely insensitive to the temperature of the heating source. We used a blackbody with a luminosity of 5$\times$10$^{40}$~erg\,s$^{-1}$ and a temperature of 5000~K for all models. \citet{fox2000} found that the X-ray luminosity of SN~1995N was about 10$^{41}$~erg\,s$^{-1}$ in 1998 and had increased by a factor of two between August 1997 and January 1998; \citet{zampieri2005} reported that in 2004, the X-ray luminosity was five times higher than the total reprocessed IR/optical flux.

\begin{figure}
	\includegraphics[width=0.45\textwidth]{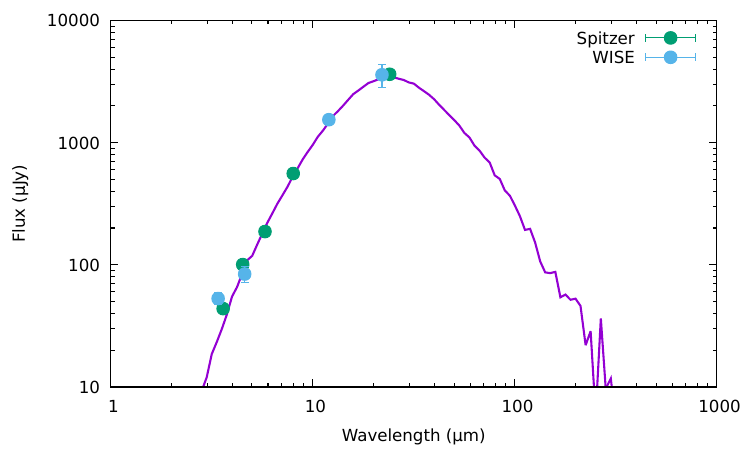}
	\caption{The best-fitting SED predicted by models of ejecta dust after 5366 days. This model has 0.4\,M$_\odot$ of amorphous carbon dust in the form of 1$\mu$m grains, distributed in a clumpy shell with a volume filling factor of 0.02.}
	\label{bestfittingSED}
\end{figure}

\begin{figure*}
\includegraphics[width=0.33\textwidth]{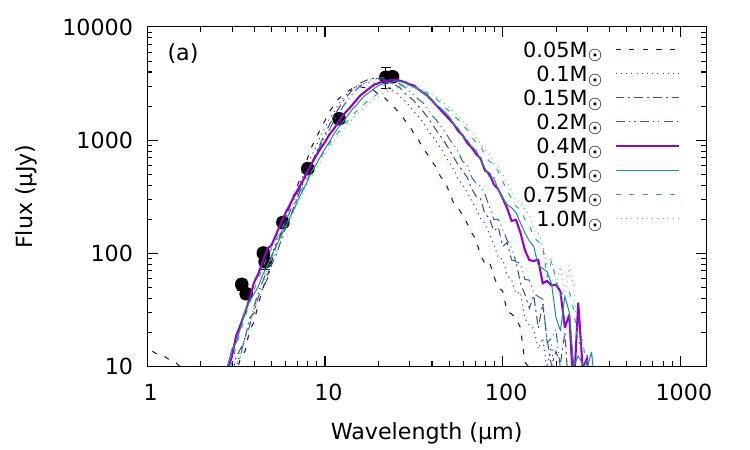}
\includegraphics[width=0.33\textwidth]{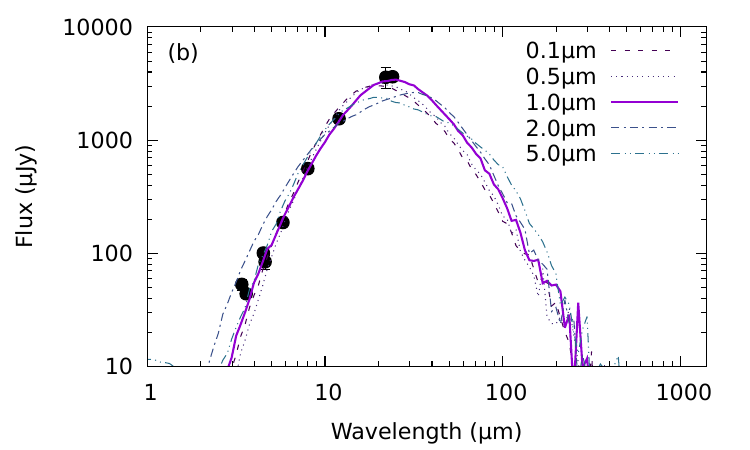}
\includegraphics[width=0.33\textwidth]{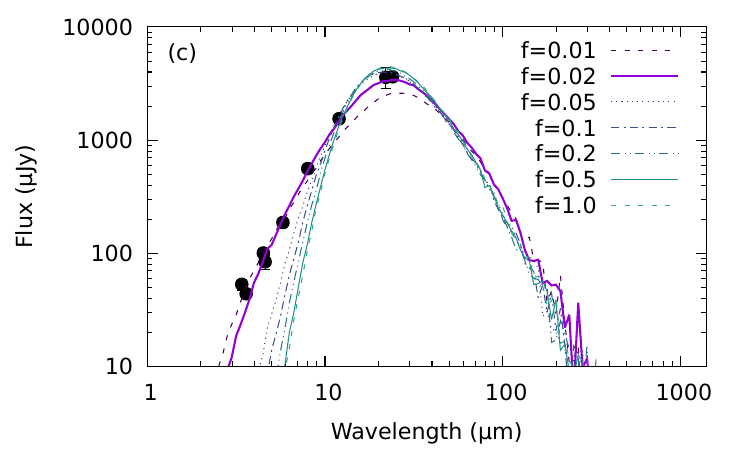}
\caption{The effect of varying model parameters on the predicted SEDs for ejecta dust heated indirectly by ejecta-CSM interaction. In each figure, the best fitting model with 0.4~M$_\odot$ of dust in clumps with a volume filling factor of 0.02, and a grain size of 1.0~$\mu$m, is shown with a solid line. (a) the effect of varying the dust mass from 0.05 to 1.0~M$_\odot$. (b) the effect of varying the grain size, from 0.1 to 5.0~$\mu$m. (c) the effect of varying the clump filling factor, from 0.01 to 1.0.}
\label{mocassinfits}
\end{figure*}

Based on this setup, we find that shells with low dust masses are heated to higher temperatures than observed and cannot reproduce the longer wavelength fluxes. At least 0.1\,M$_\odot$ of dust must be present to reproduce the Spitzer 24$\mu$m datapoint. For dust masses higher than this, the grain size, filling factor and dust mass are somewhat degenerate with each other, and similar fits can be obtained with different combinations of their values. However, we find that a model with a dust mass of 0.4~M$_\odot$ of amorphous carbon dust in the form of 1.0~$\mu$m grains, in clumps with a volume filling factor of 0.02, gives the formal best fit to the entire \textit{Spitzer}+\textit{WISE} SED. This best-fitting SED is shown in Figure~\ref{bestfittingSED}. The effect of varying dust mass, grain size and filling factor is illustrated in Figure~\ref{mocassinfits}, which shows the effect on the SED of varying each of the dust mass, grain size, and filling factor while holding the other two parameters constant. The properties of the clumps in the best-fitting model are summarised in Table~\ref{clumpproperties}.

\begin{table}
	\begin{tabular}{ll}
		\hline
		Property & Value \\
		\hline
		Number of clumps & $\sim$4300 \\
		Clump radius     & R$_{\rm out}/30$ \\
		Clump dust mass  & 9.3$\times$10$^{-5}$M$_\odot$ \\
		Clump density    & 1.5$\times$10$^{-8}$g/cm$^{3}$ \\
		\hline
	\end{tabular}
	\caption{Properties of the individual dust clumps in the best-fitting {\sc mocassin} model of the expanding ejecta.}
	\label{clumpproperties}
\end{table}

\section{Emission line profile modelling}
Our late-time observations of SN~1995N clearly show broad and asymmetric emission line profiles. The most likely cause of the asymmetry is dust formation in the ejecta; photons from the receding edge of the ejecta pass through a column of dust which photons from the near edge do not, on their way to the observer, resulting in attenuation of the red wing of intrinsically symmetric emission. Depending on the velocity of the ejecta at its inner and outer edges, the variation with radius of the velocity, and the ratio of inner to outer radii, the intrinsic profile can take a variety of forms. Dust absorption and scattering then modifies this profile, generally giving rise to an apparent blueshift. The shape of the emission line profile is not affected by any dust outside the remnant, which will affect the photons emitted from the approaching and receding parts of the ejecta equally. This is in contrast to SED modelling where it may be impossible to disentangle, for a distant supernova with no spatial information, the emission from flash-heated CSM and the emission from freshly-formed ejecta dust.

We use the radiative transfer code {\sc damocles} (\citealt{bevan2016,bevan2018}) to calculate emission line profiles from dust ejecta. The code predicts emission line profiles based on assumed velocity and emissivity distributions using a Monte Carlo technique to propagate energy packets. We first model the emission at early times using line profiles presented by \citet{fransson2002}, to determine when dust formation began and in what quantity. We then calculate models to fit archival spectra taken in 2010 (around the same time as the \textit{WISE} and \textit{Spitzer} data used for SED fitting), and our own spectra taken in 2016.

\subsection{Line profiles before day 2000}
\label{earlyspectra}

\begin{figure}
\includegraphics[width=0.5\textwidth]{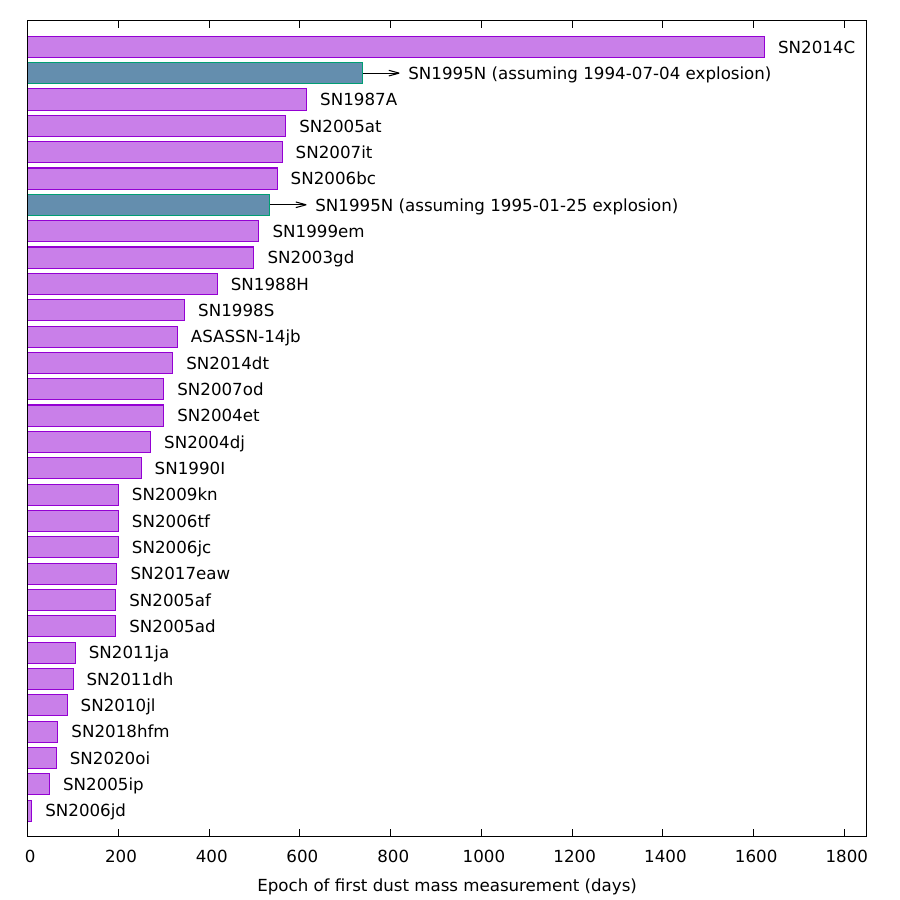}
\caption{The epoch at which the presence of freshly-formed ejecta dust is first inferred, for 28 CCSNe. For SN1995N, the epoch at which dust is first inferred is shown both for the explosion date adopted by \citet{fransson2002} and for the later date we adopt in this paper.}
\label{onset}
\end{figure}

\citet{fransson2002} presented early spectra of SN~1995N, which showed the onset of a red-blue asymmetry in the H$\alpha$ profile characteristic of dust formation. Their adoption of a July 1994 explosion date meant that their earliest spectrum was assigned an epoch of 700 days. In this spectrum, the H$\alpha$ line was close to symmetric, as it had been when the supernova was discovered. In core-collapse supernova ejecta where dust formation has been inferred, through the development of an infrared excess, a steepening optical decline, and the onset of emission line asymmetries, the epoch at which formation begins is typically much earlier. In Figure~\ref{onset}, we show the first epoch at which newly-formed dust is inferred to be present for 28 CCSNe, representing all available literature measurements as of January 2022.\footnote{The source data for this figure is available at \url{https://www.nebulousresearch.org/dustmasses}.} These epochs are subject to several caveats; pre-existing dust may be partly or wholly responsible for infrared emission at early epochs, potentially affecting some of the objects for which very early dust formation is claimed. On the other hand, sparse temporal sampling means that the first epoch at which dust is detected may come some time after the dust formation actually began. However, the median epoch of these measurements is 260 days, and only SN~2014C has its first ejecta dust mass measurement at an epoch later than 700 days; this object has dust mass estimates at epochs from 45--801 days but ejecta and CSM dust could not be distinguished (\citealt{tinyanont2016}). As discussed in Section~\ref{epoch}, although a number of statements in the literature imply that the date of SN~1995N's explosion was constrained to 10 months or more before discovery, this assumption is not well supported, and we consider a later explosion epoch of 1995 January 25 to be more likely. Adopting this explosion epoch, we model the emission line profiles presented by \citet{fransson2002}, at day 534 (their 700), 827 (their 1000) and 1589 (their 1700). We extracted the line profiles from Figure 14 of \citet{fransson2002} using WebPlotDigitizer (\citealt{rohatgi2020}).

At day 534, the H$\alpha$ line is almost symmetric. We first modelled its profile assuming no dust, and using the emissivity distribution adopted by \citet{fransson2002}: constant inside 1000~kms$^{-1}$, and proportional to V$^{-4.6}$ at higher velocities. However, we found that adjusting these parameters slightly improved the fit, and adopted 700~kms$^{-1}$ as the limiting velocity and an emissivity proportional to V$^{-4}$ in our fitting.

\begin{figure*}
\includegraphics[width=\textwidth]{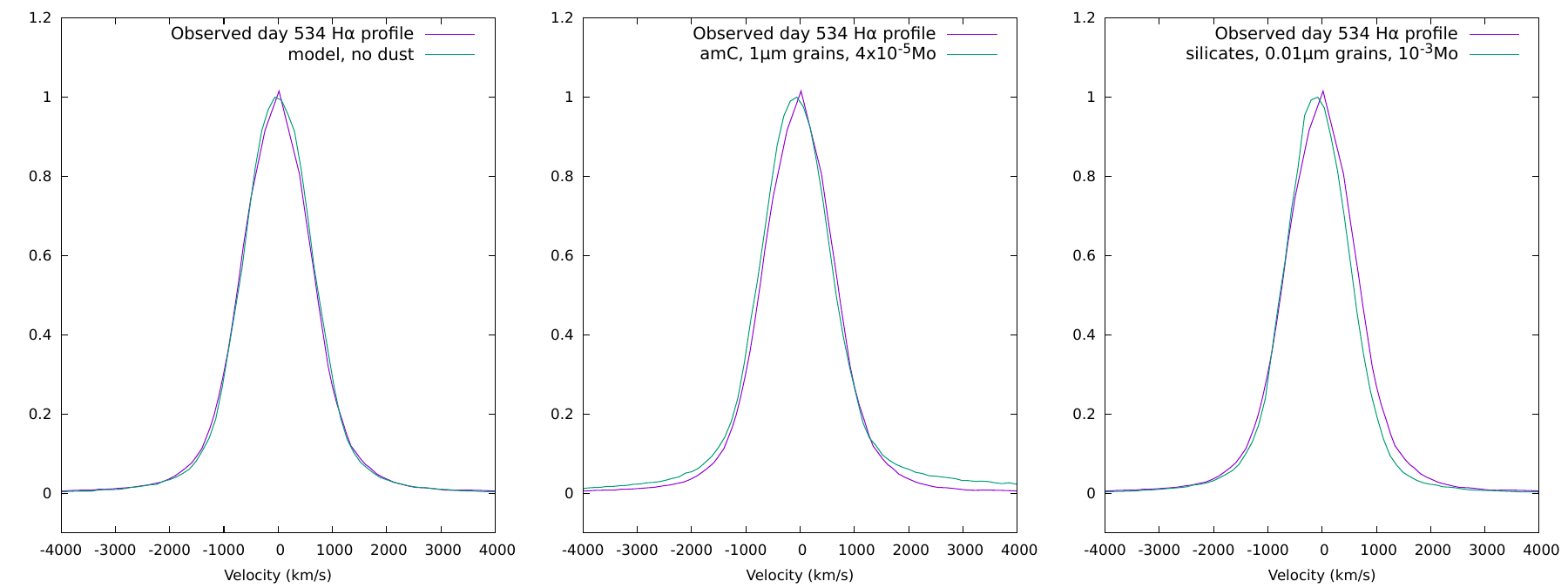}
\caption{H$\alpha$ emission line profile models at day 534. The left panel shows a model containing no dust, which fits the observed profile satisfactorily. The centre and right panels show amorphous carbon and silicate models respectively, for the dust mass at which the predicted line profile becomes inconsistent with the observations. The limiting dust masses are 4$\times$10$^{-5}$~M$_\odot$ for carbon, and 10$^{-3}$~M$_\odot$ for silicates.}
\label{d534limit}
\end{figure*}

We then examined the effect of dust on the line profile. We ran models containing either amorphous carbon dust or silicate dust, with grain sizes of 0.01, 0.1 and 1.0~$\mu$m. We find that small quantities of dust result in a profile that remains almost symmetric but is blue-shifted by a few tens of km\,s$^{-1}$. Figure 14 of \citet{fransson2002} shows the line profiles reflected about their peak and thus does not contain the necessary information to fully constrain the dust mass. However, assuming that the line profile at this epoch is not blue-shifted, we find that no more than 3$\times$10$^{-6}$~M$_\odot$ of amorphous carbon dust can be present, or 10$^{-4}$~M$_\odot$ of silicate dust. Allowing the line peak to be blue-shifted, more dust can be present before the line profile becomes inconsistent with the observations. For carbon dust, increasing the grain size allows more dust to be present and still be consistent with the observations. For silicates, however, larger grains are ruled out by the absence of a red scattering wing in the observed line profiles. We find upper limits of 4$\times$10$^{-5}$~M$_\odot$ of carbon dust, assuming 1.0~$\mu$m grains, and 10$^{-3}$~M$_\odot$ of silicate dust, assuming 0.01~$\mu$m grains. These fits are shown in Figure~\ref{d534limit}.

We then extended this model to subsequent epochs. \citet{fransson2002} do not state that the emission line peaks shifted bluewards, although their Figure 13 indicates that the peak at their day 1799 was blueshifted relative to the peak at their day 1007. \citet{bevan2016} find that dust always results in a blue-shifted emission line peak. To estimate the dust mass at these epochs, we vary the dust mass, and then shift the observed line profile to match the modelled peak velocity. At day 827, we find that for both silicates and carbon dust, only 0.01~$\mu$m grains can fit the data; as with the day 534 fits, larger grains result in a red scattering wing which is inconsistent with the observations. With 0.01~$\mu$m grains, 6$\times$10$^{-5}$~M$_\odot$ of amorphous carbon or 5$\times$10$^{-3}$~M$_\odot$ of silicates can fit the observed line profile, with blue-shifts of 160 and 230 km\,s$^{-1}$ respectively.

\begin{figure*}
\includegraphics[width=\textwidth]{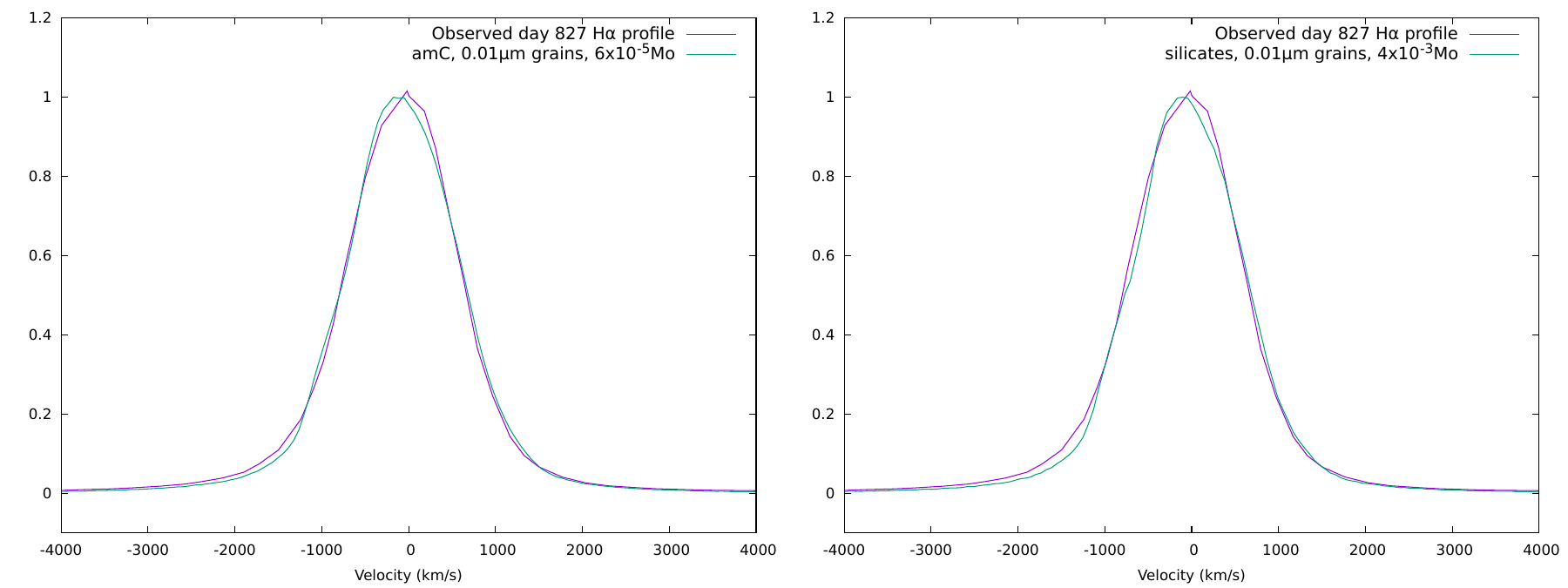}
\includegraphics[width=\textwidth]{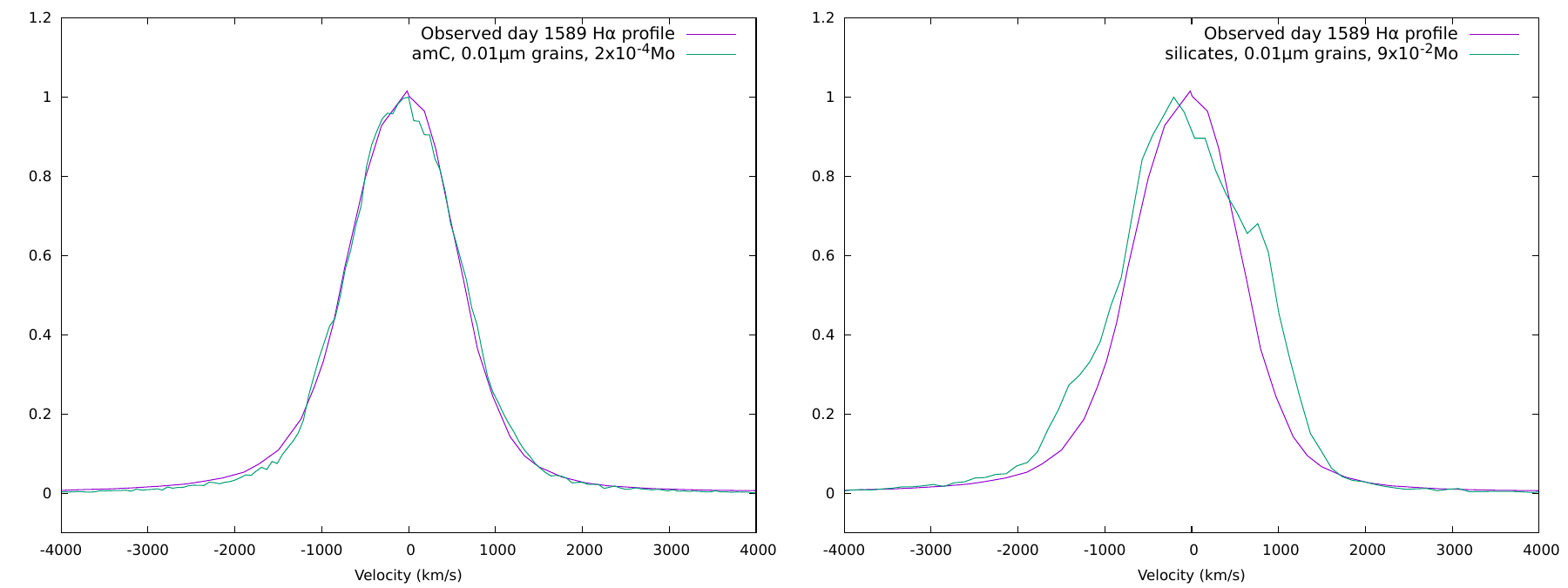}
\caption{Best-fitting H$\alpha$ emission line profile models at days 827 (top) and 1589 (bottom), for amorphous carbon (l) and silicates (r). At day 1589, silicates cannot provide a good fit for any dust mass or grain size that we considered.}
\label{d827mass}
\end{figure*}

At day 1529, the absence of a red scattering wing again constrains the grains to small sizes, and at this epoch, silicates cannot give a line profile consistent with the observations for any grain size; the modelled profile is too broad and its shape becomes irreconcilable with the smooth observed profile. For amorphous carbon, however, 2$\times$10$^{-4}$~M$_\odot$ of dust with a grain size of 0.01~$\mu$m gives a reasonable fit, when the observations are shifted by -150~km\,s$^{-1}$. We therefore conclude that the evolution of the H$\alpha$ line profile at this epoch is consistent with the gradual formation of amorphous carbon dust in the form of 0.01~$\mu$m grains. Our model fits at days 827 and 1589 are shown in Figure~\ref{d827mass}.

\subsection{Line profiles after day 5000}
\label{latespectra}
The increasing asymmetry of the H$\alpha$ line profile at early epochs provides strong evidence for the onset of dust formation in the expanding supernova ejecta after a few hundred days, supporting our interpretation of the late time infrared emission as arising from ejecta dust. From the late time emission line profiles, we have independent constraints on the dust properties. In both the 2010 archival X-shooter spectra and our own X-shooter data from 2016, broad emission lines are clearly detected. The most prominent are of [O~{\sc i}] (6300,6363), [O~{\sc ii}] (7320,7330) and [O~{\sc iii}] (4959, 5007). Hydrogen Balmer lines are also present but weak. As the [O~{\sc ii}] lines consist of two pairs of two lines, while the [O~{\sc i}] and [O~{\sc iii}] features both comprise one pair of lines, we modelled only the [O~{\sc i}] and [O~{\sc iii}] features. We assumed intrinsic line ratios of $\frac{5007}{4959}$=3.01 (\citealt{wesson2016}) and $\frac{6300}{6363}$=2.98 (\citealt{storey2000}).

We first investigated models to fit the day 5648 X-shooter spectra, which are roughly contemporaneous with the infrared photometry presented in Section~\ref{spitzerwise}. We ran {\sc damocles} models for a grid covering the same parameter space as our {\sc mocassin} models; that is, a homogeneously expanding shell of dust, with inner and outer radii corresponding to 1000 and 5000~km\,s$^{-1}$; dust masses from 0.05 to 1.0~M$_\odot$, grain radii from 0.1 to 5.0~$\mu$m, and volume filling factors for the clumpy dust distribution from 0.01 to the smooth dust case of 1.0. Although the early-time line profiles could not be reproduced with silicate dust, we ran models at the later epochs for both amorphous carbon and silicate dust.

\begin{figure*}
\includegraphics[width=\textwidth]{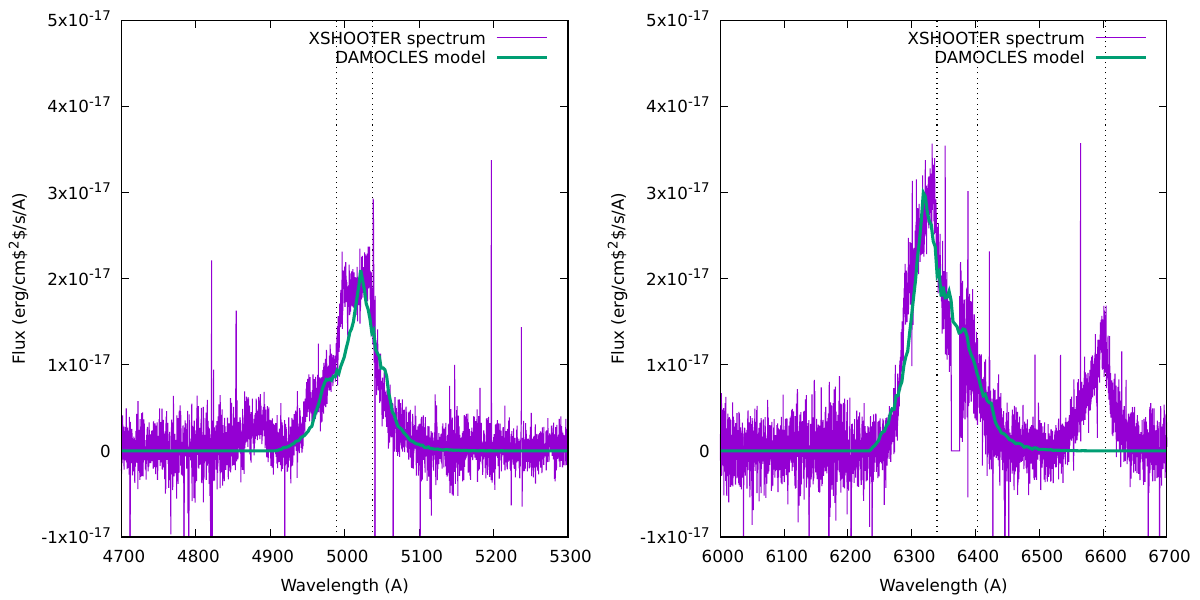}
\caption{{\sc damocles} fits to emission lines of [O~{\sc i}] (left) and [O~{\sc iii}] (right) 5648 days after the explosion. The dust configuration has a dust mass of 0.4~M$_\odot$, a grain size of 1.0~$\mu$m, and a volume filling factor of 0.05. Vertical dashed lines indicate the rest wavelengths of the emission lines at a redshift of 0.00617.}
\label{d5648fits}
\end{figure*}

We find from this grid that, as with the {\sc mocassin} models, a small volume filling factor is favoured. The smooth dust case is incompatible with the observed profiles, predicting much greater absorption of the red wing than is observed, except for 5~$\mu$m grains which produce an extended red scattering wing. Models with a filling factor of 0.01-0.02 do not predict enough absorption compared to the observations for any dust parameters. Filling factors of 0.05-0.1, though, are able to satisfactorily reproduce the profiles of both the [O~{\sc iiii}] 4959,5007 lines, and the [O~{\sc i}] 6300,6363 lines (Figure~\ref{d5648fits}). Models in this range of filling factors are almost insensitive to the mass, grain size and species of the dust. A similar result was found for day 800 observations of SN~1987A (\citealt{wesson2021}), for which filling factors of around 0.1 result in line profiles which become insensitive to the dust mass. For the largest grain sizes, the models predict a noticeable red scattering wing which is not seen in the observations. The line profiles are thus consistent with the dust properties inferred from the {\sc mocassin} modelling, and independently constrain the filling factor to small value, although they do not constrain the dust mass or grain species.



\begin{figure*}
\includegraphics[width=\textwidth]{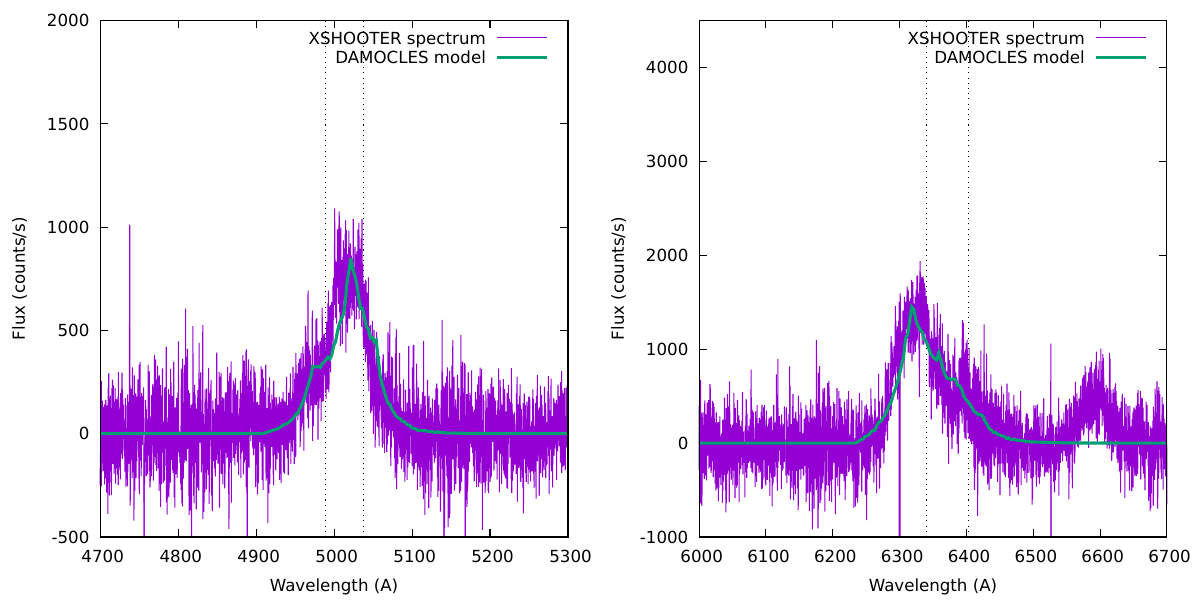}
\caption{{\sc damocles} fits to emission lines of [O~{\sc i}] (left) and [O~{\sc iii}] (right) 7858 days after the explosion. The dust configuration has a dust mass of 0.4~M$_\odot$, a grain size of 1.0~$\mu$m, and a volume filling factor of 0.05. Vertical dashed lines indicate the rest wavelengths of the emission lines at a redshift of 0.00617.}
\label{latedust}
\end{figure*}

Running the same grid of models for the day 7858 spectra, we find that similarly, the volume filling factor is the parameter to which the line profiles are most sensitive. The smooth dust case and densely filled clumpy shells result in line profiles which show either too much absorption, or a strong red scattering wing. The smallest filling factors again do not produce enough absorption to match the profiles, and the best-fitting volume filling factor is around 0.05-0.1. Once again this filling factor is in the range for which other dust properties have little effect, and the properties derived from the {\sc mocassin} models can satisfactorily reproduce the line profiles at this epoch also. The best-fitting day 7858 model fits to the [O~{\sc iii}] and [O~{\sc i}] profiles are shown in Figure~\ref{latedust}.


\section{Discussion}


We have shown that late-time infrared emission from SN~1995N is not likely to be a thermal echo from circumstellar dust, as it would require the CSM to be too massive and too far from the progenitor. However, the IR emission can be reproduced by a model in which newly-formed dust in the ejecta is heated by the ejecta-CSM interaction, which also gives rise to the X-ray emission observed from the supernova at late times. We propose that the dust heating is not direct, but rather that the X-rays from the ejecta-CSM interaction ionise gas in the ejecta, which in turn heats the dust. This model can successfully reproduce the observed dust temperatures, whereas a model in which the dust is heated directly by photons from the CSM interaction cannot. However, a lack of observational constraints on the spectrum illuminating the gas and the composition and location of the gas within the ejecta mean that our model is necessarily simplified. More detailed modelling would be required to confirm that the heating mechanism we propose is viable.

In common with a number of other studies of supernova ejecta dust (e.g. \citealt{gall2014,wesson2015,owen2015,bevan2016,priestley2019}) we find that at late times, relatively large dust grains, with radii greater than 0.1~$\mu$m, are required to fit the observations. Dust grains of this size may survive the passage of the reverse shock and contribute to the dust budget of galaxies (\citealt{kirchschlager2019}). Our best-fitting model parameters at each epoch are given in Tables~\ref{summarytable1} (early epochs) and \ref{summarytable2} (late epochs).

\begin{table*}
\begin{tabular}{lllll}
\hline
Epoch (days) & Dust species & Mass (M$_\odot$) & Grain size ($\mu$m) & Peak velocity (km\,s$^{-1}$) \\
\hline
534 & Amorphous carbon & $<$4$\times$10$^{-5}$ & $<$1.0$\mu$m  & 0 \\
534 & Silicates        & $<$10$^{-3}$         & 0.01$\mu$m & 0 \\
827 & Amorphous carbon & 6$\times$10$^{-5}$   & 0.01$\mu$m & -160 \\
827 & Silicates        & 5$\times$10$^{-3}$   & 0.01$\mu$m & -230 \\
1589 & Amorphous carbon & 2$\times$10$^{-4}$   & 0.01$\mu$m & -150 \\
\hline
\end{tabular}
\caption{The best-fitting parameters for {\sc damocles} models which reproduce emission-line profiles at epochs $<$2000 days. In all models, the emissivity is constant inside V=1000\,kms${-1}$ and $\propto$V$^{-4}$ outside.}
\label{summarytable1}
\end{table*}

\begin{table*}
\begin{tabular}{llllllll}
\hline
Epoch (days) & Technique & Inner and outer radii (km\,s$^{-1}$) & Dust species & Mass (M$_\odot$) & Grain size ($\mu$m) & Volume filling factor \\
\hline
5366 & SED                    & 1000--5000 & Amorphous carbon & 0.4    & 1.0$\mu$m & 0.02 \\
5648 & Emission line profiles & 1000--5000 & Amorphous carbon & $>$0.1 &           & 0.05 \\
7858 & Emission line profiles & 1000--5000 & Amorphous carbon & $>$0.1 &           & 0.05 \\
\hline
\end{tabular}
\caption{The best-fitting parameters for {\sc mocassin} and {\sc damocles} models for the SED and emission line profiles in 2009--2010 and 2017.}
\label{summarytable2}
\end{table*}


The developing asymmetries in emission line profiles in early data can be reproduced with the onset of dust formation after day $\sim$500; by day $\sim$1600, the dust mass is limited to less than a few $\times$10$^{-4}$~M$_\odot$, and the grain size is constrained to radii of $\sim$0.01~$\mu$m. Emission line profiles at days 5648 and 7858, meanwhile, independently support the clumpy dust geometry with a small volume filling factor derived from {\sc mocassin} models, though they do not strongly constrain other dust properties. Between $\sim$4 and $\sim$15 years after the supernova explosion, therefore, we find that the dust mass increased by a factor of $\sim$100. This implies continuing dust formation long after theoretical models indicate that it should have ceased (\citealt{sarangi2015,sluder2018}). \citet{dwek2015} argued that this tension between observations and theory can be resolved if the dust forms rapidly at early times in extremely optically thick clumps. In the case of SN~1987A, this was ruled out by \citet{wesson2015}, who found from radiative transfer models that the clump filling factor in the ejecta was constrained by the amount of optical emission escaping, and was incompatible with clumps dense enough to conceal large quantities of dust. They gave an upper limit of 0.001~M$_\odot$ of dust at 1015 days after the explosion. Independently, \citet{bevan2016} used emission line profiles to determine the dust mass in the ejecta at epochs up to 3000 days, also finding that large dust masses at early times were not compatible with the observations. \citet{wesson2021} reinforce these earlier conclusions by considering the SED and emission line profiles simultaneously, finding that while a carefully-chosen clump filling factor can result in line profiles which are insensitive to the dust mass, the SED is still sensitive to it. Likewise, a model with a high dust mass which does reproduce the SED cannot also reproduce the line profiles. In the case of SN~1995N, the gradual onset of asymmetry in the profile of H$\alpha$ between 500 and 1500 days after the explosion also rules out the presence of a large mass of dust at this epoch.

Further observational evidence of dust forming over many years in supernova ejecta comes from dust grains found in meteorites. \citet{liu2018} found a positive correlation between $^{49}$Ti and $^{28}$Si excesses in presolar silicon carbide (SiC) dust grains from supernovae. Attributing this to the radioactive decay of $^{49}$V to $^{49}$Ti, they inferred that the SiC grains had condensed at least two years after the supernova explosion. Similarly, \citet{ott2019}, also looking at SiC grains, found that the isotopic composition of barium in such grains was consistent with them having formed after a significant fraction of freshly produced $^{137}$Cs had already decayed into $^{137}$Ba. This decay has a half-life of 30 yr, implying that the grains condensed about 20 yr after the supernova explosion. While models predict that SiC should precipitate from the gas later than the species which comprise the majority of the dust mass (\citealt{sarangi2015,sluder2018}), these isotopic results show that dust can continue to form on decadal timescales, rather than being completed within a few years of the supernova explosion.

\citet{bevan2020} found that for SN~2010jl, some specific dust configurations can be found in which the emission line profile becomes insensitive to the dust mass. However, they found that in these configurations, the SED remains sensitive to the total dust mass, which is thus not unconstrained. We find this to be the case in SN~1995N as well. Radiative transfer models of the SED require a dust mass of $\sim$0.4~M$_\odot$ in clumps with a volume filling factor of $\sim$0.02, and for such small filling factors, the emission line profiles become insensitive to the dust mass and grain size. Nevertheless, they are consistent with the substantial dust mass inferred from the SED, and models of early-time emission line profiles show that such a large mass could not have been present before day 1500.

\begin{figure*}
\includegraphics[width=0.75\textwidth]{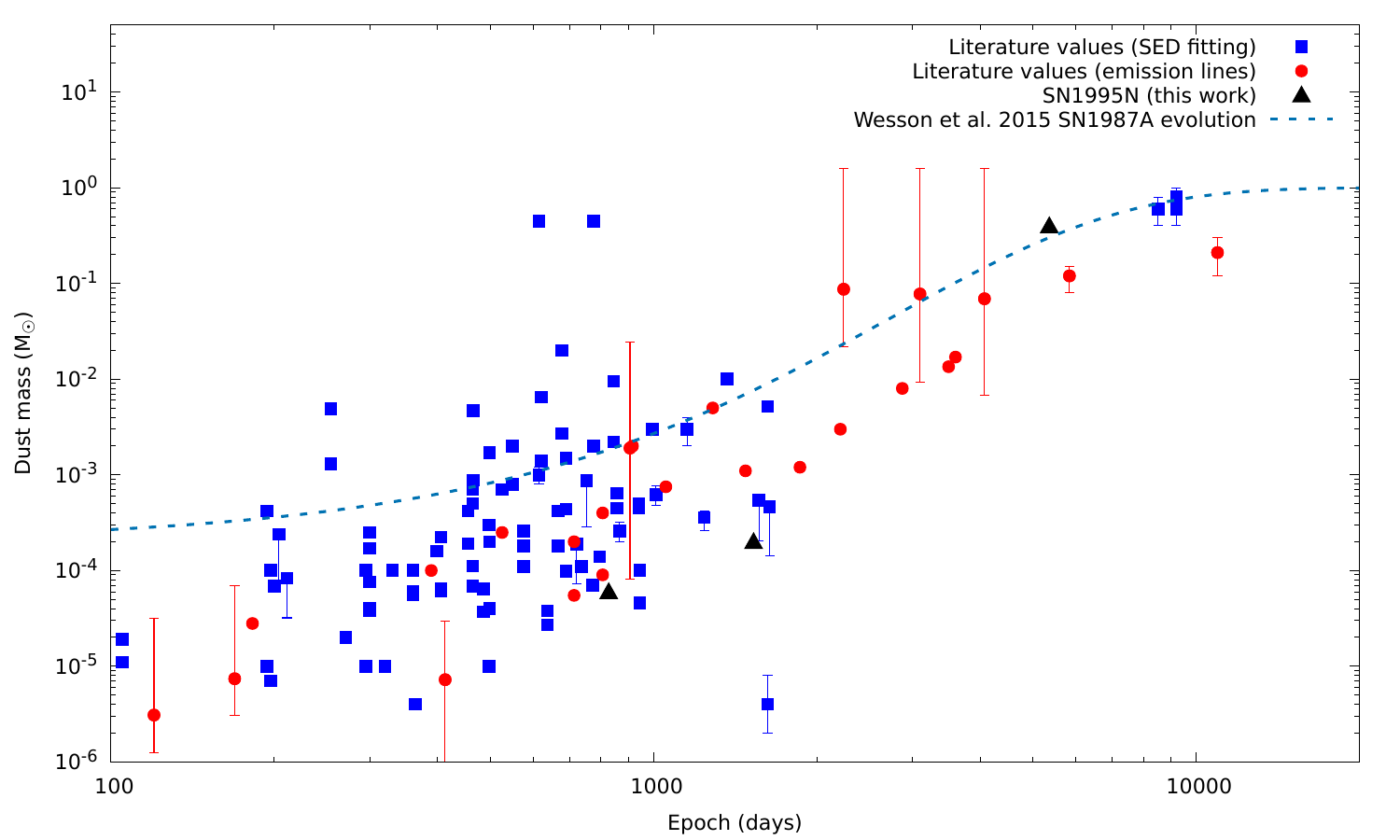}
\caption{A compilation of all available literature values for supernova ejecta dust masses, with the masses found in the current work for SN~1995N indicated. The sigmoid curve proposed by \citet{wesson2015} to fit the observed evolution of the dust mass of SN~1987A is shown as a dashed grey line. The literature data used to create this figure is available at \url{https://www.nebulousresearch.org/dustmasses}.}
\label{newsigmoid}
\end{figure*}

The pattern of dust growth that we derive for SN~1995N is consistent with the overall pattern of observational estimates of dust masses with time. In Figure~\ref{newsigmoid}, we show a compilation of all published supernova ejecta dust masses, with the values derived in this work indicated. Although estimates at early times span several orders of magnitude, and estimates at very late times remain relatively sparse, the overall pattern of slow increases in observationally-inferred dust masses, reaching cosmologically-significant values of 0.1--1~M$_\odot$ only at epochs $\gg$1,000 days, is apparent. As we have only three data points for the dust mass in the ejecta of SN~1995N, we cannot meaningfully fit the simple three-coefficient sigmoid equation used by \citet{wesson2015} to describe the evolution of the dust mass in SN~1987A. However, the early dust masses in SN~1995N are at the lower end of the range found for all CCSNe. Figure~\ref{newsigmoid} shows points colour-coded according to the means by which they were estimated; it can be seen that in general, dust masses derived from emission line profiles are at the lower end of the overall range. As the shape of emission line profiles can only be affected by dust within the emitting region, and is insensitive to any circumstellar material outside the emitting region, this may suggest that at these early epochs, SED-based dust mass estimated might be systematically overestimating dust masses due to the presence of CSM dust. However, the small number of objects with emission line-based dust mass estimates, and the very small number of objects with contemporaneous determinations from both techniques, means that this possible systematic difference between emission-line and SED-based dust mass estimates cannot yet be confirmed.

With the ever-increasing evidence that large masses of dust do form in many supernova remnants, the major question is how much of this dust will survive the passage of the reverse shock and go on to contribute to the dust budget of galaxies. The dust grain size is a critical factor in this. We have found that, while early-time spectra can only be fitted with very small grains (a=0.01~$\mu$m), the late-time SED and line profiles require micron-sized grains. \citet{kirchschlager2019} found that the survival of carbon grains was maximised for grain radii of 0.5-1.5~$\mu$m. They found that the optimal radius for grain survival depends on the density contrast between dust clumps and the interclump medium, with higher contrasts resulting in the survival of smaller dust grains. A number of studies have now found that micron-sized dust grains are required to fit late-time observations of supernova ejecta dust; these include studies of SN~2010jl (\citealt{gall2014}), the Crab Nebula (\citealt{temim2013, owen2015}), and SN~1987A, in which \citet{wesson2015} found that dust grains were predominantly micron-sized by 23 years after the explosion, while the SED at 600-800 days could not be well fitted with such large grains.

We thus conclude that SN~1995N exhibits the following properties, in common with a number of other supernovae:
\begin{enumerate}[i]
\item a mass of newly formed dust of the order of several tenths of a solar mass, a quantity sufficient to account for the dust masses found in some high-redshift quasars (\citealt{morgan2003,dwek2007};
\item evidence for continuing dust formation many years after the supernova, in contrast to current theoretical models of dust formation; and
\item evidence for dust grains exceeding 0.1~$\mu$m in radius, indicating that a significant fraction of the dust could survive the passage of the reverse shock.
\end{enumerate}

Such properties have now been found both in normal Type II supernovae, and in Type IIn supernovae which have shown interaction of the ejecta with the CSM.

\section{Acknowledgments}
Based on observations collected at the European Organisation for Astronomical Research in the Southern Hemisphere under ESO programmes  084.D-0265(A) and 097.D-05(A).
This work is based in part on archival data obtained with the Spitzer Space Telescope, which was operated by the Jet Propulsion Laboratory, California Institute of Technology under a contract with NASA. Support for this work was provided by an award issued by JPL/Caltech.
This publication makes use of data products from the Wide-field Infrared Survey Explorer, which is a joint project of the University of California, Los Angeles, and the Jet Propulsion Laboratory/California Institute of Technology, funded by the National Aeronautics and Space Administration.

RW, AB and MJB are supported by European Research Council (ERC) Grant SNDUST 694520. IDL acknowledges support from ERC grant DustOrigin 851622.
MM  acknowledges support from STFC Ernest Rutherford fellowship (ST/L003597/1) and STFC Consolidated grant (2422911).

\section{Data availability}

The raw X-shooter data analysed in this work are available from the ESO archive facility at \url{http://archive.eso.org/}.

\bibliographystyle{mnras}
\bibliography{references} 

\end{document}